\documentclass[final,1p,times]{elsarticle}

\usepackage{amssymb}
\usepackage{hyperref}

\journal{Nuclear Physics B}
\usepackage{amsfonts}
\usepackage{amsthm}
\usepackage{amsmath}
\newcommand{\dd}{\text{d}}

\newcommand{\ee}{\text{e}}

\newcommand{\p}{\partial}

\newcommand{\eps}{\varepsilon}

\newcommand{\sW}{{\text{\tiny W}}}
\usepackage{color}

\usepackage{tikz}
\usepackage{circuitikz}
\usetikzlibrary{arrows,shapes,positioning}
\usetikzlibrary{arrows,shapes,positioning}
\usetikzlibrary{decorations.markings}
\tikzstyle arrowstyle=[scale=1]
\tikzstyle directed=[postaction={decorate,decoration={markings,
    mark=at position .65 with {\arrow[arrowstyle]{stealth}}}}]
\tikzstyle reverse directed=[postaction={decorate,decoration={markings,
    mark=at position .65 with {\arrowreversed[arrowstyle]{stealth};}}}]
\usepackage[percent]{overpic}

\begin{document}

\begin{frontmatter}

\title{Weakly universal dynamical correlations between eigenvalues of large random matrices}

\author[first]{Kirone Mallick}
\affiliation[first]{organization={Institut de Physique Théorique, CEA Saclay \& CNRS (UMR 3681)},%Department and Organization
            addressline={}, 
            city={Gif-sur-Yvette},
            postcode={91190},
            country={France}}
\author[second]{Gabriel Téllez}
\affiliation[second]{organization={Departamento de Física, Universidad de los Andes},%Department and Organization
           % addressline={}, 
            city={Bogotá},
            postcode={ 111711},
            country={Colombia}}
            
\author[third,fourth]{Frédéric van Wijland}  
\affiliation[third]{organization={Laboratoire Mati\`ere et Syst\`emes Complexes, Université Paris Cité  \& CNRS (UMR 7057)},%Department and Organization
            %addressline={}, 
            city={Paris},
            postcode={75013}, 
            country={France}}
\affiliation[fourth]{organization={Yukawa Institute for Theoretical Physics, Kyoto University},%Department and Organization
            %addressline={,
%Kitashirakawaoiwake-cho, Sakyo-ku}, 
            city={Kyoto},
            postcode={606-8502},
            country={Japan}}

\begin{abstract}
It was shown roughly thirty years ago that the density correlations of eigenvalues of large random matrices display a universal form, independent of most of the details of the distribution of the random matrix itself. We show that when the matrix elements evolve according to a Dyson Brownian motion, dynamical correlations retain a large degree of the universality found at equal times when expressed in terms of the characteristics of some partial differential equation in the complex plane.
\end{abstract}

\begin{keyword}
%% keywords here, in the form: keyword \sep keyword, up to a maximum of 6 keywords
random matrices \sep macroscopic fluctuation theory \sep hydrodynamics 

\end{keyword}

\end{frontmatter}

\section{Random matrices and hydrodynamics}
\label{sec:intro}
It is impossible to do justice in a short introduction to the seventy year old history of random matrices. When Wigner~\cite{wigner1993characteristic} suggested the spectral properties of the nuclei of heavy atoms could be similar to those of matrices with random independent entries, this drew the interest of the theoretical and mathematics physics community. The long range Coulomb gas nature of the mutual interactions between eigenvalues drew Dyson's interest who endowed the matrix elements with a Brownian motion dynamics~\cite{dyson1962brownian} of their own. We refer the reader to the excellent textbooks covering random matrices and their statistical properties~\cite{mehta2004random,forrester2010log,anderson2010introduction,livan2018introduction,potters2020first}. Statistical dynamical aspects, arising from Dyson Brownian motion, are a very lively area of research. In pure mathematics, $p-$adic versions have very recently been built~\cite{castro2003p,van2023p}; Dyson Brownian motion has also been shown to be instrumental in  the study of high dimensional statistics~\cite{bun2016application,bodin2024random} (see also the references contained in these two recent theses); in theoretical physics we refer the reader to active versions~\cite{touzo2023interacting} or to a variation with resetting~\cite{biroli2025resetting}. Our work is perhaps closer in spirit to recent endeavors in probability theory, such as in~\cite{huang2019rigidity} (see \cite{Guionnet2009} for a recent review of the mathematical approach to Dyson Brownian motion): we are interested in the dynamics of the fluid of eigenvalues for large random matrices. After all, there are only very few strongly interacting systems whose dynamics can be solved exactly, and this is of course a very interesting aspect of the Dyson Brownian motion in our view.\\

Following~\cite{brezin1978planar,t1993planar}, we are using a distribution for the $N\times N$ random matrix $M$ of the form $\ee^{-\beta N \text{Tr}(V(M))}$, where $M$ belongs to either of the standard ensembles (the parameter $\beta$ is adjusted so that $\beta=1$ for the GOE, $\beta=2$ for the GUE and $\beta=4$ for the GSE). We restrict our analysis to $V$ being an even polynomial of degree $n$ in the matrix $M$ and such that the distribution of eigenvalues eventually stays confined to a symmetric connected interval (the sufficient criteria can be found in \cite{eynard2015random}) and can be written as $\mu_\sW(\lambda)=Q(\lambda)\sqrt{a^2-\lambda^2}$, where $Q$ is a $V$-dependent polynomial of degree $n-2$ and $a$ is a $V$-dependent constant. For large random matrices in the GUE ensemble, Brézin and Zee ~\cite{brezin1993universality, brezin1993universal} established that when appropriately smoothed, the correlation function of the eigenvalue distribution takes a universal form in equilibrium, namely a form that is largely independent of the potential $V$ entering the distribution of the matrix elements. Their result was soon after extended by Beenakker to all standard ensembles~\cite{BEENAKKER1994515} by a method circumventing the orthogonal polynomials used in \cite{brezin1993universality}. Their expression for the static structure factor at $\lambda\neq\lambda'$ is
\begin{equation}\label{eq:brezinzee}
    C_\text{eq}(\lambda,\lambda')=-\frac{1}{\beta N^2\pi^2}\frac{a^2-\lambda\lambda'}{\sqrt{a^2-\lambda^2}\sqrt{a^2-\lambda'^2}(\lambda-\lambda')^2}
\end{equation}
where large $N$ is assumed. Equivalently, as shown in \ref{app:corr}, the correlations of the resolvent of the empirical eigenvalue density read
\begin{equation}\begin{split}\label{eq:correl-resolv-eqint}
    \Gamma_\text{eq}(z,z')&=-\frac{1}{\beta N^2}\frac{a^2- zz'+\sqrt{z^2-a^2}\sqrt{z'^2-a^2}}{\sqrt{z^2-a^2}\sqrt{z'^
    2-a^2}(z-z')^2}
\end{split}\end{equation}
As is visible on Eqs~\eqref{eq:brezinzee} and \eqref{eq:correl-resolv-eqint} the only dependence on the potential $V$ lies in the location of the branch cut $a$; the polynomial $Q$ does not appear.\\

The goal of this work is to determine the correlation function of the eigenvalue distribution in equilibrium but at unequal times, when the matrix elements are evolving according to a Dyson Brownian motion. Our result is that some, but not all, of the universality observed in the statics, survives the dynamics. The main result of this work is that the dynamical correlations of the resolvent are given by
\begin{equation}\begin{split}\label{eq:kgf}
    \Gamma(z,z';t)&=
    -\frac{1}{\beta N^2}\frac{Q(\zeta)}{Q(z)}\frac{a^2- \zeta z'+\sqrt{\zeta^2-a^2}\sqrt{z'^2-a^2}}{\sqrt{z^2-a^2}\sqrt{z'^
    2-a^2}(\zeta-z')^2}
\end{split}\end{equation}
where the function $\zeta(t,z)$ is a characteristic of an auxiliary dynamical system defined by
\begin{equation}
    \frac{\dd\zeta}{\dd t}=Q(\zeta)\sqrt{\zeta^2-a^2}
\end{equation}
with $\zeta(0)=z$. Note that in time-dependent correlations the polynomial $Q$ (and therefore the potential $V$) now appears explicitly.\\

The outline of this work goes as follows. In Section~\ref{sec:soa} we specify the random matrix model we study and we recall  the definition of the Dyson Brownian motion. In Section~\ref{sec:hydroresolvent}, we build a fluctuating hydrodynamics equation for the resolvent, in the spirit of the McKean-Vlasov formulation~\cite{chan1992wigner}. This allows us, in Section~\ref{sec:MFT}, to build on the macroscopic fluctuation theory~\cite{bertini2015macroscopic} to establish our main result Eq.~\eqref{eq:kgf}. In Section~\ref{sec:charact}, we translate this result on the resolvent correlations into one for the density correlations, in the specific cases of a quadratic and a quartic confining potential $V$. The conclusion gathers what we believe are stimulating research directions for the future.

\section{State of the art}\label{sec:soa}
\subsection{Statics: the universality of correlations}
For a random matrix $M$ of size $N$ belonging to one of the three standard ensembles and whose elements are distributed according to $P_\text{eq}(M)=\ee^{-\beta N \text{Tr}(V(M))}/Z$, it is well-known that the $N$ eigenvalues $\lambda_1,\ldots,\lambda_N$ are distributed according to the Boltzmann like distribution
\begin{equation}\begin{split}
    p_\text{eq}(\lambda_1,\ldots,\lambda_N)=&\frac{1}{Z(N,\beta)}\exp\Big(-\beta N\sum_i V(\lambda_i)\Big)\prod_{i<j}|\lambda_i-\lambda_j|^{\beta}\\
    = &\frac{1}{Z(N,\beta)}\exp\left[-\beta N\sum_i V(\lambda_i)+\frac{\beta}{2}\sum_{i\neq j}\ln|\lambda_i-\lambda_j|\right]
\end{split}\end{equation}
where $Z(N,\beta)$ is a normalization constant. For large $N$ they are typically a distance $1/N$ apart from each other, so that for large $N$, the empirical density 
\begin{equation}
    \hat{\mu}(\lambda)=\frac{1}{N}\sum_i\delta(\lambda-\lambda_i)
\end{equation}
converges towards a smooth function $\mu_\sW(\lambda)$. When $V(\lambda)=\frac 12 \lambda^2$ the function $\mu_\sW$ is the celebrated Wigner semi-circle distribution. The energy $N^2 E[\hat{\mu}]$ of an eigenvalue configuration can be expressed in terms of $\hat{\mu}$ as
\begin{equation}\label{eq:energy}
E[\hat{\mu}]=\int\dd\lambda V(\lambda)\hat{\mu}(\lambda)+\frac 12 \int\dd\lambda\dd\lambda' \hat{\mu}(\lambda) \ln|\lambda-\lambda'|\,\hat{\mu}(\lambda')
\end{equation}
Because entropic terms are extensive in $N$, we know that for large $N$ the most probable configuration $\mu_\sW$ is one that minimizes $E$, which leads to
\begin{equation}\label{eq:distrib}
V'(\lambda)={\mathcal P}\int\dd\lambda'    \frac{\mu_\sW(\lambda')}{\lambda-\lambda'}
\end{equation}
and this results in the generic form
\begin{equation}\label{eq:wignerQ}
    \mu_\sW(\lambda)=Q(\lambda)\sqrt{a^2-\lambda^2}
\end{equation}
where $Q$ is a polynomial of degree $n-2$ if $V$ is of degree $n$. Both the location of the branch cut $a$ and the explicit form of $Q$ are strongly $V$-dependent.  Let us quote the explicit results for two simple potentials. For $V(\lambda)=\frac{\lambda^2}{2}$ we have $Q(\lambda)=1$ and $a^2=2$, while for $V(\lambda)=\frac{\lambda^4}{4}$ we have $Q(\lambda)=\lambda^2+a^2/2$ and $a^4=\frac{8}{3}$. We refer the interested readers to \cite{brezin1978planar,brezin1993universality,zuber2012introduction} for the technical details and specific illustrations (an alternative derivation of Eq.~\eqref{eq:wignerQ} will be presented further down).\\

The static structure factor $C_\text{eq}$ is defined by
\begin{equation}
    C_\text{eq}(\lambda,\lambda')=\langle(\hat{\mu}(\lambda)-\mu_\sW(\lambda))(\hat{\mu}(\lambda')-\mu_\sW(\lambda'))\rangle
\end{equation}
This quantity, determined in \cite{brezin1993universality, BEENAKKER1994515}, is given in Eq.~\eqref{eq:brezinzee}. To derive this result, it would tempting to use the energy functional $E[\mu]$ in Eq.~\eqref{eq:energy} without too much care, but according to \cite{livan2018introduction} and to \cite{sandier20152d}, how to take into account the coinciding point behavior in this energy functional is not settled yet. For instance, one could think that the fluctuations of $\hat{\mu}$ around $\mu_\sW$ are well captured by expanding $E[\mu_\sW+\psi]$ to quadratic order around $\psi=0$, but the resulting expansion leads to $E[\mu]-E[\mu_\sW]=\frac{1}{2}\int\dd\lambda\dd\lambda'\ln|\lambda-\lambda'|\psi(\lambda)\psi(\lambda')$, in which any trace of the confining potential has been lost (and translation invariance has been restored). The careful derivation in \cite{BEENAKKER1994515}, based on a static response formalism,  rather starts from Eq.~\eqref{eq:distrib}.\\

Since we want to explore the fate of these correlations are unequal times, we must first define the dynamical evolution we will be using.\\

\subsection{Dynamics: Dyson Brownian motion}
The Dyson Brownian motion~\cite{dyson1962brownian} consists in endowing the matrix elements with an overdamped relaxational dynamics with a noise term tuned so that the stationary distribution is exactly the prescribed $P_\text{eq}(M)$:
\begin{equation}
    \frac{\dd M}{\dd t}=-\frac{\p \text{Tr} V(M)}{\p M}+\text{noise}
\end{equation}
where the noise matrix is white and Gaussian with a $1/\sqrt{\beta N}$ amplitude; its properties are ensemble dependent. When converted into an equation for the eigenvalues, this results in
\begin{equation}\label{eq:defDBM}
    \frac{\dd\lambda_i}{\dd t}=-V'(\lambda_i)+\frac{1}{N}\sum_{j\neq i}\frac{1}{\lambda_i-\lambda_j}+\sqrt{\frac{1}{\beta N}}\eta_i
\end{equation}
where the $\eta_i$'s are independent Gaussian white noises with correlations $\langle\eta_i(t)\eta_j(t')\rangle=\delta_{ij}\delta(t-t')$. Equation \eqref{eq:defDBM} is our starting point (see \cite{potters2020first} for  a pedagogical derivation).\\

By analogy to the statics, it is convenient to introduce the time-dependent empirical density 
\begin{equation}\label{eq:defempirical}
    \hat{\mu}(\lambda,t)=\frac{1}{N}\sum_i\delta(\lambda-\lambda_i(t))
\end{equation}
An overdamped dynamical evolution of the form Eq.~\eqref{eq:defDBM} generically leads the empirical density to be the solution of a stochastic partial differential equation known as the Dean-Kawasaki equation~\cite{dean1996langevin,kawasaki1994stochastic,illien2025dean}. The latter can be cast in the form
\begin{equation}\label{eq:conseq}
    \partial_t\hat{\mu}=-\p_\lambda j
\end{equation}
where the fluctuating eigenvalue current $j$ comprises a deterministic part $j_\text{det}$ and a stochastic one,
\begin{equation}
    j(\lambda,t)=j_\text{det}(\lambda,t)+\sqrt{\frac{2\hat{\mu}}{\beta N^2}}\xi(\lambda,t)
\end{equation}
where $\xi$ is a Gaussian white noise with correlations $\langle\xi(\lambda,t)\xi(\lambda',t')\rangle=\delta(\lambda-\lambda')\delta(t-t')$. The deterministic contribution $j_\text{det}$ has the explicit expression
\begin{equation}\label{eq:DK}
    j_\text{det}(\lambda,t)=
    -\frac{1}{N}\frac{2-\beta}{2\beta}\p_\lambda \hat{\mu}(\lambda,t)
    -V'(\lambda)\hat{\mu}(\lambda,t)
    +\hat{\mu}(\lambda,t){\mathcal P}\int\dd\lambda'\frac{\hat{\mu}(\lambda',t)}{\lambda-\lambda'}
\end{equation}
where the first Fick term in $j_\text{det}$ expresses that the eigenvalue diffusion is subleading at large $N$ and it will henceforth be omitted. The transport mechanism fully rests on the nonlocal interactions between eigenvalues. Pedagogical accounts of the derivation of Eq.~\eqref{eq:DK} in the context of random matrices can be found in \cite{touzo2023interacting, kirone2024}. This Dean-Kawasaki equation exactly matches the McKean-Vlasov equation obtained much earlier~\cite{chan1992wigner}) within  a more mathematical framework. As already noted by Dean~\cite{dean1996langevin}, to leading order in $N$, the conserved current related to $\hat{\mu}$ can be cast in the form
\begin{equation}
    j=-\hat{\mu}\p_\lambda\frac{\delta E}{\delta\hat{\mu}}+\sqrt{\frac{2\hat{\mu}}{\beta N^2}}\xi
\end{equation}
where $E$ is given is Eq.~\eqref{eq:energy}. At this stage, in the spirit of \cite{kirone2024}, the techniques of the Macroscopic Fluctuation Theory could in principle apply, but the confining potential competes with the logarithmic repulsion, which results in a nontrivial stationary profile $\mu_\sW$ that makes calculations difficult. A useful trick consist in considering the fluctuating Stieltjes transform of the empirical density $\hat{\mu}$ and in working with that transform.

\section{Noisy Burgers equation for the resolvent}\label{sec:hydroresolvent}
The equation satisfied by $\hat{\mu}$ can be converted into an equation for the so-called resolvent
\begin{equation}
    \hat{G}(z,t)=\int\dd\lambda\frac{\hat{\mu}(\lambda,t)}{z-\lambda}
\end{equation}
Inverting the Stieltjes transform is done by means of the Sokhotski–Plemelj theorem, which leads to the Stieltjes-Perron formula, namely
\begin{equation}\label{eq:PerronStieltjes}\hat{\mu}(\lambda,t)=\frac{1}{2i\pi }(G(\lambda-i\eps,t)-G(\lambda+i\eps,t))=\frac{1}{\pi}\text{Im}\,\hat{G}(\lambda-i\eps,t)
\end{equation}
as $\eps\to 0^+$ (because $\hat{\mu}$ is real valued).  At the deterministic level for a quadratic potential (or none), when translating this equation into one for the resolvent, it is known that the Burgers equation shows up. We shall now see how noise can be incorporated. Applying the Stieltjes transform to Eq.~\eqref{eq:conseq} leads to
\begin{equation}
    \p_t\hat{G}=-\p_z J
\end{equation}
where $J(z,t)=\int\dd\lambda\frac{j(\lambda,t)}{z-\lambda}$ is the Stieltjes transform of the physical current. As is well-known~\cite{Forrester_Grela_2016, warchol2014dynamic} (and references therein), the log-repulsion term in $j$ transforms into  a Burgers contribution:
\begin{equation}
    \int\dd\lambda\frac{1}{z-\lambda}\hat{\mu}(\lambda,t){\mathcal P}\int\dd\lambda'\frac{\hat{\mu}(\lambda',t)}{\lambda-\lambda'}=\frac 12 \hat{G}(z,t)^2
\end{equation}
The other two contributions  which we now discuss, are, to the best of our knowledge, new.\\

Let's begin with the noise term. Since $\xi(\lambda,t)$ is Gaussian and white, so is the linear combination \begin{equation}
    \Xi(z,t)=\int\dd\lambda\sqrt{\frac{2\hat{\mu}(\lambda,t)}{\beta N^2}}\frac{\xi(\lambda,t)}{z-\lambda}
\end{equation}
and it is fully determined by its correlations
\begin{equation}\label{eq:bruit}\begin{split}
    \langle\Xi(z,t)\Xi(z',t')\rangle=&\int\dd\lambda\dd\lambda'\sqrt{\frac{2\hat{\mu}(\lambda,t)}{\beta N^2}}\sqrt{\frac{2\hat{\mu}(\lambda',t')}{\beta N^2}}\frac{1}{(z-\lambda)(z'-\lambda')}\langle\xi(\lambda,t)\xi(\lambda',t')\rangle\\
    =&\frac{2}{\beta N^2}\delta(t-t')\int\dd\lambda\frac{\hat{\mu}(\lambda,t)}{(z-\lambda)(z'-\lambda')}\\
    =&-\frac{2}{\beta N^2}\delta(t-t')\p_z\p_{z'}\frac{\hat{G}(z,t)-\hat{G}(z',t)}{z-z'}
\end{split}\end{equation}
This shows that the noise on the resolvent is white in time but correlated in space.\\

The other piece in the current we have to transform is the $-V'(\lambda)\hat{\mu}(\lambda,t)$ contribution. For an arbitrary potential, this is a nontrivial task. At this stage it is convenient to recall that once $\hat{G}$ is known, the moments of $\hat{\mu}$ can in principle be accessed. We denote by $\hat{m}_k(t)=\int\dd\lambda \lambda^k\hat{\mu}(\lambda,t)$ the $k^{\rm th}$ empirical moment. This quantity is also the $k^{\rm th}$ coefficient of the Laurent expansion of the resolvent at $z=\infty$. We introduce the fluctuating quantity $\hat{\gamma}_k(z,t)=\int\dd\lambda\frac{\lambda^k\hat{\mu}(\lambda,t)}{z-\lambda}$. It  is such that $ \hat{m}_k(t)=\lim_{z\to\infty}z\gamma_k(z,t)$. Thanks to the recursion relation
\begin{equation}
    \gamma_k(z,t)=z \gamma_{k-1}-\hat{m}_{k-1}
\end{equation}
and using that $\gamma_0=\hat{G}$, we see that
\begin{equation}
    \hat{\gamma}_k(z,t)=z^k\hat{G}(z,t)-\sum_{j=0}^{k-1} z^{k-1-j}\hat{m}_j(t)
\end{equation}
so that the moments are given recursively by
\begin{equation}
    \hat{m}_k(t)=\lim_{z\to\infty}z(z^k\hat{G}(z,t)-\sum_{j=0}^{k-1} z^{k-1-j}\hat{m}_j(t))
\end{equation}
We are now in a position to determine the Stieltjes transform of $V'(\lambda)\hat{\mu}(\lambda,t)$:
\begin{equation}
 \int\dd\lambda\frac{1}{z-\lambda} V'(\lambda)\hat{\mu}(\lambda,t)=V'(z)\hat{G}(z,t)-\hat{P}(z,t)  
\end{equation}
where $\hat{P}$ is a polynomial in $z$ with degree $n-2$
when $V$ is of degree $n$. Its coefficients are linear combinations of the empirical moments $\hat{m}_j(t)$, such that $V'(z)\hat{G}(z,t)-\hat{P}(z,t)$ decays as $O(1/z)$ as $z\to\infty$. It is therefore still a fluctuating quantity.\\

For a quadratic potential $V(\lambda)=\frac{\lambda^2}{2}$, $\gamma_1(z,t)=z\hat{G}-1$ and thus $\hat{P}(z,t)=1$ is rather trivial. But for a quartic potential $V(\lambda)=\frac{\lambda^4}{4}$, we find that 
\begin{equation}
    \hat{P}(z,t)=z^2+\hat{m}_1(t) z+\hat{m}_2(t)
\end{equation}
and for $V(\lambda)=\frac{\lambda^6}{6}$ we have
\begin{equation}
    \hat{P}(z,t)=z^4+\hat{m}_1 z^3+\hat{m}_2 z^2+\hat{m}_3 z+\hat{m}_4
\end{equation}
We should keep in mind that $P$ is itself a fluctuating observable that is linear in $\hat{G}$ (or $\hat{\mu}$). This is a polynomial of degree $n-2$ (where $n$ is the degree of $V$) for which the coefficient of $z^{n-2}$ is that of $z^{n-1}$ in $V'$.\\

To conclude we have shown that the resolvent $\hat{G}$ evolves according to the following conservation equation
\begin{equation}\label{eq:DKG}
    \p_t\hat{G}=-\p_z J
\end{equation}
where the current $J$ has the expression
\begin{equation}\label{eq:DKG2}
    J=-V'(z)\hat{G}+\frac 12 \hat{G}^2+\hat{P}+\Xi
\end{equation}
where the Gaussian noise $\Xi$ has correlations given by
\begin{equation}\label{eq:bruitjoli}
\langle\Xi(z,t)\Xi(z',t')\rangle=-\frac{2}{\beta N^2}\delta(t-t')\p_z\p_{z'}\frac{\hat{G}(z,t)-\hat{G}(z',t)}{z-z'}
\end{equation}
as derived in Eq.~\eqref{eq:bruit}.
The price to pay for converting the nonlocal part of the current in eigenvalue space into a local function of $\hat{G}$ is the emergence of a nonlocal noise. The $\hat{G}^2$ contribution is a Burgers nonlinearity. When the noise term can be omitted, and for the specific case of a quadratic potential, this observation has been exploited many times, see for instance \cite{PhysRevE.82.051115,warchol2014dynamic} and references therein. Equation \eqref{eq:DKG} is thus a McKean-Vlasov like equation for the resolvent. It expresses in the abstract space of the $z$ variable conjugate to the eigenvalues, the existence of a fluctuating hydrodynamics with a weak noise in the form of a noisy Burgers equation of a new type. This prompts us to implement the method of the Macroscopic Fluctuation Theory on $\hat{G}$, rather than on $\hat{\mu}$.\\

As a side remark, we would like to draw the reader's attention to the fact that the resolvent method can also be used to recover the eigenvalue distribution beyond the quadratic potential case. At large $N$ and in equilibrium the empirical resolvent converges to a fixed function $G_\sW(z)$. In order to find $G_\sW$ one may simply impose the mean current to vanish,
\begin{equation}
    \langle J\rangle=0=-V'(z)G_\sW(z)+\frac 12 G_\sW^2+P_\sW(z)
\end{equation}
where $P_\sW(z)$ is a polynomial of degree $n-2$ (with leading coefficient that of $V'$, and whose other coefficients are linear combinations of the moments of $\mu_\sW$). This immediately tells us that 
\begin{equation}\label{eq:solresol}
    G_\sW(z)=V'(z)-\sqrt{V'(z)^2-2P_\sW(z)}
\end{equation}
Asking for a single symmetric branch cut (because $V$ is even) forces a factorization of the form
\begin{equation}\label{eq:solresol2}
    G_\sW(z)=V'(z)-Q(z)\sqrt{z^2-a^2}
\end{equation}
where $Q$ is a polynomial of degree $n-2$ and $a$ a constant. We emphasize that the polynomials $Q$ and $P$, and the constant $a$, are uniquely fixed by comparing the leading terms in the large $z$ behavior of Eqs.~\eqref{eq:solresol} and \eqref{eq:solresol2}. The first terms in the expansion determine $Q$, then $a$, and finally, pushing the identification to higher orders in $1/z$ allows one to find $P_\sW$. Of course, inverting the Stieltjes transform readily brings us back to the eigenvalue distribution Eq.~\eqref{eq:wignerQ}. The resolvent (and the algebraic-only manipulations that come with it) can thus be used for an arbitrary potential, which is extensively discussed in \cite{eynard2015random} (and references therein), even for potentials leading to a non connected support.

\section{Dynamical correlations of the resolvent}\label{sec:MFT}
\subsection{Macroscopic Fluctuation Theory for the resolvent}
We find it convenient to formulate the dynamics of Eqs.~\eqref{eq:DKG} and \eqref{eq:DKG2} in terms of the Janssen-De Dominicis action governing the weight of a path $G$ of the resolvent and of the associated response field $\bar{G}$, which is exponential in the large parameter $\beta N^2$, namely $\ee^{-\beta N^2 S[\bar{G},G]}$, where the action reads
\begin{equation}\label{eq:JDDResolvent}\begin{split}
    S[\bar{G},G]=&\int\dd t\dd z\bar{G}\left(\p_t G+\p_z \left[- V'(z){G}+P+\frac{1}{2}{G}^2\right]\right)\\
    &-\int\dd t\dd z \dd z'\p_z \bar{G}(z,t)\p_{z'}\bar{G}(z',t)(-1)\frac{{G}(z,t)-{G}(z',t)}{z-z'}
\end{split}\end{equation}
While the results presented in this work could be derived by working directly at the level of the stochastic partial differential equation for $\hat{G}$, we use the Janssen-De Dominicis path integral formalism with later applications to large deviations in mind (for which this formalism is essential). The expression in Eq.~\eqref{eq:JDDResolvent} is the starting point of the subsequent analysis.\\ 

Our ultimate interest goes to the dynamical structure factor
\begin{equation}
    C(\lambda,\lambda';t-t')=\langle\hat{\mu}(\lambda,t)\hat{\mu}(\lambda',t')\rangle-\mu_\sW(\lambda)\mu_\sW(\lambda')
\end{equation}
First we shall consider the corresponding correlations in $z$ space, which we denote by
\begin{equation}
    \Gamma(z,z';t-t')=\langle{G}(z,t){G}(z',t')\rangle-G_\sW(z)G_\sW(z')
\end{equation}
where the angular brackets refer to an average with respect to the $\ee^{-\beta N^2 S[\bar{G},G]}$ weight.\\

Since the noise amplitude decays as $\frac{1}{\beta N^2}$ the correlation function $\Gamma$ will be given, to leading order in $N$, by the action $S$ truncated to quadratic order by writing $G=G_\sW+\phi$ and $\bar{G}=\bar{\phi}$. Keeping in mind that $P(z)$ is linear in $G$, its coefficients can be split into a deterministic and a random contribution: $m_k=m_{k,\sW}+\delta m_k$ where $m_{k,\sW}$ is the $k^{\rm th}$ moment of $\mu_\sW$ and 
\begin{equation}
    \delta m_k=\lim_{Z\to\infty}\left[Z^{k+1}\phi(Z,t)-\sum_{j=1}^{k-1}Z^{k-j}\delta m_j\right]
\end{equation}
for $k\geq 1$. These $\delta m_k$ terms will contribute to the action at the quadratic order through $\delta P(z)=P(z)-P_\sW(z)=\sum_{k=1}^{n-2}\delta m_k z^{n-2-k}$ (note that $\delta m_0=0$). Thus
the quadratic action reads
   \begin{equation}\begin{split}\label{eq:JDDlinquad}
    S[\bar{\phi},\phi]=&\int\dd t\dd z\bar{\phi}\left(\p_t\phi+\p_z(-(V'-G_\sW)\phi+\delta P(z)\right)\\
    &+ \int\dd t\dd z\dd z'\bar{\phi}(z,t)\p_z\p_{z'}\frac{G_\sW(z)-G_\sW(z')}{z-z'}\bar{\phi}(z',t)
\end{split}\end{equation} 
We are ultimately after $\Gamma(z,z';t-t')=\langle\phi(z,t)\phi(z',t')\rangle$. Standard methods~\cite{tauber2014critical} tell us that an intermediate step before obtaining $\Gamma$ is to solve for the response function $R(z,z';t-t')=\langle\phi(z,t)\bar{\phi}(z',t')\rangle$ which is a solution of
\begin{equation}\label{eq:forRgeneral}
    \p_t R(z,z';t-t')-\p_z[Q(z)\sqrt{z^2-a^2}\,R(z,z';t)-P_R(z;z',t-t')]=\delta(t-t')\delta(z-z')
\end{equation}
where $P_R(z;z',t-t')=\langle\delta P(z,t)\bar{\phi}(z',t')\rangle$ is a polynomial in $z$ of degree $n-2$. This is a linear functional of $R$ built in such a way that $V'(z)R(z,z';t-t')-P_R(z;z',t-t')$ decays as $1/z$ when $z\to\infty$. Its coefficients are functions of $z'$ and $t-t'$.
Then, once $R$ is determined, one solves for $\Gamma$:
\begin{equation}\label{eq:forGamma}\begin{split}
    \p_s \Gamma(z,z';s-t')-\p_z[Q(z)\sqrt{z^2-a^2}\,\Gamma(z,z';s-t')-P_\Gamma(z;z',s-t')]=\\-\frac{2}{\beta N^2}\int\dd z''\p_{z}\p_{z''}\frac{G_\sW(z)-G_\sW(z'')}{z-z''}R(z',z'';t'-s)
\end{split}\end{equation}
where $P_\Gamma(z;z',s-t')$ is a polynomial in $z$ (with $z'$ and $s-t'$ dependent coefficients) such that $V'(z)\Gamma(z,z';s-t')-P_\Gamma(z;z',s-t')$ decays as $1/z$ when $z\to\infty$. We solve and discuss the solution of Eqs.~\eqref{eq:forRgeneral} and \eqref{eq:forGamma} in the next two subsections. We defer to subsection \ref{subsec:char} and to  \ref{subsec:technical} a brief discussion of the counterterms introduced by $P_R$ and $P_\Gamma$.\\

Before closing this subsection, let's consider,  as an example, the quartic potential $V(\lambda)=\frac{\lambda^4}{4}$, for which $P(z)=z^2+m_1 z+m_{2}$, with $P_\sW(z)=z^2+\sqrt{8/27}$ and  $\delta P(z)=z\delta m_1+\delta m_2$ with
\begin{equation}
   \delta m_1=\lim_{Z\to\infty}Z^2\phi(Z,t),\;\;\delta m_2= \lim_{Z\to\infty}(Z^3\phi(Z,t)-Z\delta m_1)
\end{equation}
In that case, the equation for $\Gamma$ becomes
\begin{equation}\begin{split}
    \p_s \Gamma(z,z';s-t')-\p_z[Q(z)\sqrt{z^2-a^2}\,\Gamma(z,z';s-t')]+\lim_{Z\to\infty} Z^2\Gamma(Z,z';s-t')=\\-\frac{2}{\beta N^2}\int\dd z''\p_{z}\p_{z''}\frac{G_\sW(z)-G_\sW(z'')}{z-z''}R(z',z'';t'-s)
\end{split}\end{equation}

\subsection{The method of characteristics for the response}\label{subsec:char}
The equation \eqref{eq:forRgeneral} determining the response function would be easier to solve without the contribution that is involving the counterterms $P_R(z)$ that depend on the asymptotic behavior of $R$ at infinity  (regardless of the explicit expression of $V$, this contribution takes the form of a polynomial in $z$). Temporarily forgetting this contribution allows Eq.~\eqref{eq:forRgeneral} to be solved by the method of characteristics. We shall consider the characteristic $\zeta(t)$ run backwards in time, such that $\zeta(0)=z$, which evolves according to 
\begin{equation}\label{eq:charact}
    \frac{\dd\zeta}{\dd t}=Q(\zeta)\sqrt{ \zeta^2-a^2}
\end{equation}
Using the $\delta(z-z')$ initial condition for $R$ leads to
\begin{equation}\label{eq:solforR}
    R(z,z';t)=\theta(t)\frac{Q(\zeta(t))\sqrt{\zeta(t)^2-a^2}}{Q(z)\sqrt{z^2-a^2}}\delta(\zeta(t)-z')=\theta(t)\frac{Q(z')\sqrt{z'^2-a^2}}{Q(z)\sqrt{z^2-a^2}}\delta(\zeta(t)-z'))
\end{equation}
and $\delta(\zeta(t,z)-z')$ can also be rewritten as \begin{equation}\label{eq:solforRaux}
    R(z,z';t)=\theta(t)\frac{1}{Q(z)\sqrt{z^2-a^2}}\delta(t-\tau(z',z))
\end{equation} 
where we have used that 
 $Q(z')\sqrt{z'^2-a^2}    \delta(\zeta(t,z)-z')=\delta(t-\tau(z',z))$,  $\tau(z',z)$ being the time taken by the characteristic to go from $z'$ to $z$.\\

It will later prove convenient to introduce an auxiliary set of variables $\xi(t)$, $\chi$ and $\chi'$ such that 
\begin{eqnarray}\label{eq:deffauxangles}
    \zeta(t)=a\cosh\xi(t),\;\;z=a\cosh\chi,\;\; z'=a\cosh\chi'
\end{eqnarray}
along with the function $q$ defined by $q(\xi)=Q(\zeta)$. In terms of these quantities the characteristic and the response function become
\begin{equation}\label{eq:caractxi}
    \frac{\dd\xi}{\dd t}=q(\xi),\,\,\xi(0)=\chi
    \end{equation}
and
\begin{equation}
    R(z,z';t)=\theta(t)\frac{q(\xi(t))}{a\sinh\chi'q(\chi)}\delta(\xi(t)-\chi')
\end{equation}
The asymptotic behavior of the characteristics depends on the potential. For a quadratic potential ($n=2$), that is $V(\lambda)=\frac{\lambda^2}{2}$ and $Q(\lambda)=1$, we readily obtain
\begin{equation}\label{eq:solcharactquad}
    \zeta(t)=z\cosh t+\sqrt{z^2-2}\sinh t
\end{equation}
and as $t\to+\infty$ we see that $\zeta(t)\to\infty$. In other words, it takes an infinite time to reach infinity. By contrast, when the potential is of higher degree ($n>2$), the phenomenology of the characteristic considerably differs: the characteristic equation \eqref{eq:charact} exhibits a finite time blow up, that is, it now takes a finite time to go from $z'$ to infinity, $\tau(z',\infty)<+\infty$.\\

We are now in a position to analyze the role of the polynomial $P_R$ appearing in Eq.~\eqref{eq:forRgeneral}. In the quadratic potential case, $P_R=0$, and thus $R$ given in Eq.~\eqref{eq:solforR} is the full solution of Eq.~\eqref{eq:forRgeneral}. For higher order potential, at fixed time $t>0$, we see that $R(z,z';t)$ in Eq.~\eqref{eq:solforRaux} decreases as $\frac{1}{z Q(z)}\propto \frac{1}{z^{n-1}}$ as $z\to\infty$, because $\zeta(t,z)\to \zeta(t,\infty)$ remains finite for $t>0$. Therefore, at $t>0$, we have $\lim_{Z\to\infty} Z^k R(Z,z';t)=0$ for $k=0,\ldots,n-2$. The polynomial $P_R$ is thus independent of $z$ and is given by
\begin{equation}\label{eq:PR}
    P_R(z;z',t)=Q(z')\sqrt{z'^2-a^2} \delta(\zeta(t,\infty)-z')
\end{equation}
Since  $\p_z P_R(z;z',t)=0$ for any $t>0$, the extra terms in Eq.~\eqref{eq:forRgeneral}  self-consistently vanish. To conclude, the function $R$ given by Eq.~\eqref{eq:solforRaux} is indeed the correct solution for an arbitrary potential.

%Il me reste le doute de ce qui se passe a $t=0$ car cet argument ne marche plus et il me semble qu'il reste des contre termes non nuls quand $t=0$.
%En effet, serait-il même possible que cela contribue en $\delta(t-t')$ ? Je suggère que l'on n'en dise pas davantage, car au fond on a déjà le résultat qui nous intéresse.

%\begin{figure}
%	\centering 
%	\includegraphics[width=0.4\textwidth, angle=-90]{Nuclear_Physics_B_cover_image.pdf}	
%	\caption{Nuclear Physics B journal cover} 
%	\label{fig_mom0}%
%\end{figure}

\subsection{Exact calculation of the correlations}
We now investigate the solution of Eq.~\eqref{eq:forGamma}. The response function being causal, we see that for $s>t'$, the equation loses its right-hand side and it can, again, be solved by the method of characteristics, provided the additional terms $P_\Gamma$ involving limits at infinity are temporarily discarded. Fortunately we know the equal time expression of $\Gamma$, given by
\begin{equation}
    \Gamma_\text{eq}(z,z')=\Gamma(z,z';0)
\end{equation}
because it is the double Stieltjes transform of $C_\text{eq}(\lambda,\lambda')$ given in Eq.~\eqref{eq:brezinzee} (and found in \cite{brezin1993universality,BEENAKKER1994515}), so that the initial condition of the characteristic is known. It is shown in \ref{app:corr} that
\begin{equation}\begin{split}\label{eq:correl-resolv-eq}
    \Gamma_\text{eq}(z,z')
&=-\frac{1}{\beta N^2}\frac{a^2- zz'+\sqrt{z^2-a^2}\sqrt{z'^2-a^2}}{\sqrt{z^2-a^2}\sqrt{z'^
    2-a^2}(z-z')^2}\\
    &=\frac{1}{2\beta N^2 a^2}
    \frac{1}{\sinh\chi\sinh\chi'\sinh^2\frac{\chi+\chi'}{2}}
\end{split}\end{equation}
where $\chi$ and $\chi'$ are defined in Eq.~\eqref{eq:deffauxangles},
and thus
\begin{equation}\begin{split}\label{eq:kgf}
    \Gamma(z,z';t)=&
    -\frac{1}{\beta N^2}\frac{Q(\zeta)}{Q(z)}\frac{a^2- \zeta z'+\sqrt{\zeta^2-a^2}\sqrt{z'^2-a^2}}{\sqrt{z^2-a^2}\sqrt{z'^
    2-a^2}(\zeta-z')^2}\\
    =&
    \frac{1}{2\beta N^2 a^2}
    \frac{q(\xi)}{q(\chi)}
    \frac{1}{\sinh\chi\sinh\chi'
    \sinh^2\frac{\xi(t)+\chi'}{2}}
\end{split}\end{equation}
It is now time to return to the full equation comprising the terms involving the limit of $\Gamma$ at infinity and using the discussion as above Eq.~\eqref{eq:PR} we see again that as $z\to\infty$ and $t>0$ the term involving $P_\Gamma$ in Eq.~\eqref{eq:forGamma} does not contribute.\\

Equation \eqref{eq:kgf} is the central result of this work. It shows that when expressed in terms of the characteristic trajectory $\zeta(t)$, the dynamical correlations of the fluctuating resolvent retain a large degree of universality. Of course, the potential $V$ appears through the explicit expressions of $Q$ and $a$, and of the full function $\zeta(t)$. In that sense, one could also view this result as a clear deviation from universality, since $\zeta(t)$ heavily depends on the details of $V$.

\section{The quadratic and the quartic potentials}\label{sec:charact}
In this section we use the general result to derive explicit expressions in the $V(\lambda)=\frac{\lambda^2}{2}$ and $V(\lambda)=\frac{\lambda^4}{4}$ cases, and we derive the large time asymptotics of the eigenvalue density correlations.

\subsection{The harmonic case}
In general, it is a nontrivial task to invert the Stieltjes transform, simply because this requires the explicit knowledge of the characteristic $\zeta(t)$ and of its $z$ dependence. However, in the angular-like variables defined by $\zeta=a\cosh\xi$, the characteristic equation \eqref{eq:charact} simplifies into
\begin{equation}
    \label{eq:charact-angular}
    \frac{\dd\xi}{\dd t}=1.
\end{equation}
and then $\xi(t)=t+\chi$. Direct substitution into Eq.~\eqref{eq:kgf} and using $a^2=2$ leads to
\begin{equation}
\label{eq:Gamma-pot2}
    \Gamma(z,z';t)=
    \frac{1}{4\beta N^2 }
    \frac{1}{\sinh\chi\sinh\chi'\sinh^2\frac{t+\chi+\chi'}{2}}.
\end{equation}
Its long time decay as $t\to+\infty$ is given by
\begin{equation}
    \Gamma(z,z';t)\simeq
    \frac{1}{\beta N^2 }
    \frac{\ee^{-\chi}}{\sinh\chi}
    \frac{\ee^{-\chi'}}{ \sinh\chi'} \ee^{-t}.
\end{equation}
To return to the eigenvalue space $\lambda=a\cos\theta$, one has to compute the discontinuity of $\Gamma$ from $\chi=-i\theta$ to $\chi=i\theta$ for both $\chi$ and $\chi'$. For the long time behavior, this is easy to do, because the contributions from $\chi$ and $\chi'$ are factorized, leading to 
\begin{align}
    C(\lambda,\lambda';t)&
    \simeq\frac{1}{\pi^2\beta N^2 }
    \frac{\cos\theta}{\sin\theta}
    \frac{\cos\theta'}{\sin\theta'}
    \ee^{-t}
    \\
    &\simeq\frac{1}{\pi^2\beta N^2 }\frac{\lambda\lambda'}{\sqrt{2-\lambda^2}\sqrt{2-\lambda'^2}}\ee^{-t}
\end{align}
It is interesting to note that while the equal time correlations are negative, thus expressing the well-known Coulomb repulsion, for a large time separation, fluctuations are actually positively correlated.\\ 

For the full expression Eq.~\eqref{eq:Gamma-pot2} at arbitrary time, the passage to the eigenvalue space yields
\begin{equation}
   C(\lambda,\lambda';t)=
   \frac{1}{4\pi\beta N^2}
   \frac{1}{\sin\theta\sin\theta'}
   \text{Re}
   \left[ 
   \frac{1}{\sinh^2\frac{t+i(\theta+\theta')}{2}}
   +
   \frac{1}{\sinh^2\frac{t+i(\theta-\theta')}{2}}
   \right]
   \end{equation}
When converted into an expression in terms of the eigenvalues, this results in
\begin{equation}
    C(\lambda,\lambda';t)=
    \frac{-4-4\lambda^2\lambda'^2+\lambda\lambda'(7+2(\lambda^2 +\lambda'^2))\cosh(t)
    -4(\lambda^2+\lambda'^2-1)\cosh(2t))+\lambda\lambda'\cosh(3t)}%
    {2\pi^2\beta N^2 \sqrt{(2-\lambda^2)(2-\lambda'^2)}
    \left[\lambda^2+\lambda'^2-1-2\lambda\lambda'
    \cosh(t)+\cosh(2t)\right]^2}
\end{equation}
which is valid for $t\geq 0$.

\subsection{For a quartic potential and beyond}\label{subsec:quarticcorrel}

\begin{figure}[h]
    \centering
  %  \includegraphics[width=10cm]{flowquartic.pdf}
  % \begin{overpic}[width=10cm,grid,tics=10]{flowquartic.pdf}
   \begin{overpic}[width=10cm]{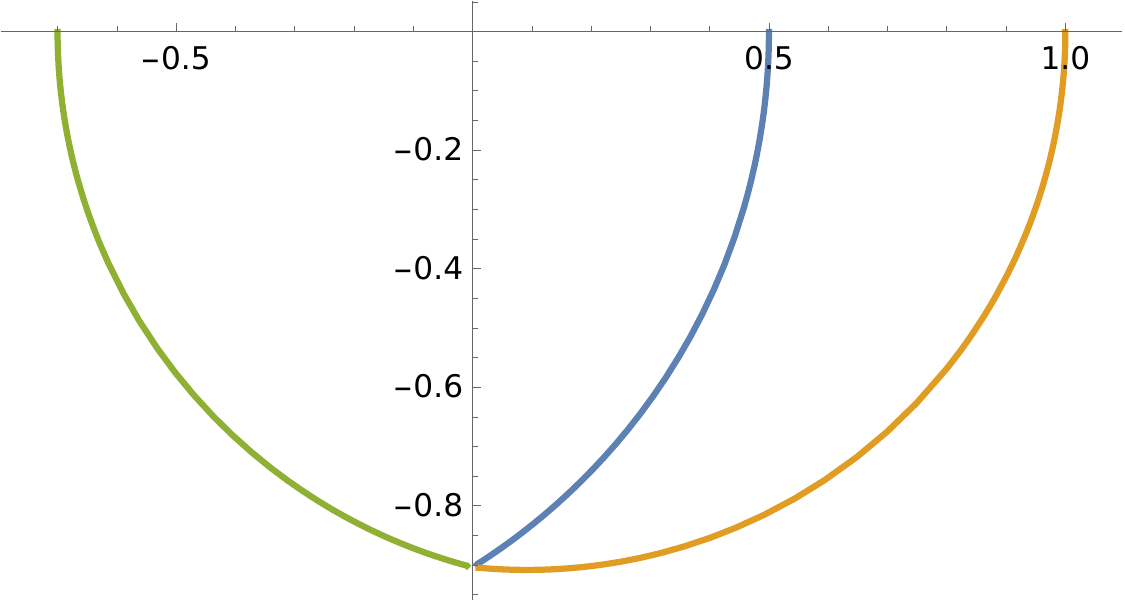}
 \put  (36,-2) {$z_1=-ia/\sqrt{2}$}
 \put (40,55) {$\text{Im}(\zeta)$}
\put (100,50) {$\text{Re}(\zeta)$}
\end{overpic}

\caption{Trajectories of the characteristics starting from right below the branch cut $[-a,a]$ for a quartic potential $V(\lambda)=\frac{\lambda^4}{4}$ ($Q(\lambda)=\lambda^2+a^2/2$ and $a^2=2\sqrt{2/3}$). All flow to the $z_1=-ia/\sqrt{2}$ root of $Q$ lying in the lower half plane (the initial values are $-0.7$ (green), $0.5$ (blue) and $1$ (orange)).}\label{fig:flotquartic}
\end{figure}

For a quartic potential $V(\lambda)=\frac{\lambda^4}{4}$ we have $Q(z)=z^2+a^2/2$ (with $a^2=2\sqrt{2/3}$), which, for the purpose of our discussion, we rewrite as $Q(z)=z^2-z_1^2$, with $z_1=-ia/\sqrt{2}$. Solving Eq.~\eqref{eq:charact} at short times leads to
\begin{equation}
  \zeta(t)=  \lambda-i0^+-iQ(\lambda)\sqrt{a^2-\lambda^2} t+\frac 12 Q(\lambda)(\lambda Q(\lambda)-(a^2-\lambda^2)Q'(\lambda))t^2+O(t^3)
    \end{equation}
where we have chosen the initial value $z=\lambda-i0^+$, with $\lambda\in[-a,a]$ (this will be required to perform the inverse Stieltjes transform). We see that, when starting right below the real axis, the short time correction brings the trajectory closer to the vertical axis (the correction has a sign opposite to that of $\lambda$) and drives it to the only available root of $Q$, $z_1=-ia/\sqrt{2}$ in the lower half plane. The flow in the complex plane is illustrated in Fig.~\ref{fig:flotquartic}.
The derivation of the approach to $z_1$ can be quantified more precisely. The analytics turn out to be much simpler using the angle like variables $\chi$, $\xi$ and $\chi_1$ such that $\cosh\chi=z/a$, $\cosh\chi_1=z_1/a$, and $\cosh\xi(t)=\zeta(t)/a$. In terms of these variables, $\xi(t)$ now flows to  $\chi_1=\ln\frac{1+\sqrt{3}}{\sqrt{2}}-i\frac{\pi}{2}$. The polynomial $Q$ becomes
\begin{equation}
    q(\xi)=a^2 \sinh(\xi+\chi_1)\sinh(\xi-\chi_1)
\end{equation}
The explicit integration of Eq.~\eqref{eq:caractxi} leads to 
\begin{equation}
\ee^{-a^2\sqrt{3} t}=\frac{\sinh(\xi(t)-\chi_1)\sinh(\chi+\chi_1)}{\sinh(\xi(t)+\chi_1)\sinh(\chi-\chi_1)}
\end{equation}
where we have used that $\xi(0)=\chi$. This leads to the following large time asymptotics 
\begin{equation}\label{eq:qsurq}
    \xi(t)-\chi_1\simeq-\sqrt{3}\ee^{-\sqrt{3} a^2 t}\frac{\sinh(\chi-\chi_1)}{\sinh(\chi+\chi_1)}
\end{equation}
so that
\begin{equation}\label{eq:qsurq}
   \frac{q(\xi)}{q(\chi)}\simeq 3\ee^{-\sqrt{3} a^2 t}\frac{1}{\sinh^2(\chi+\chi_1)}
\end{equation}
When the flow is initialized with $z$ slightly above the real axis, the characteristic $\zeta$ flows to $-z_1$ ($\xi$ to $\chi_1^*)$ and similar formulas are obtained.\\

We start from the exact Eq.~\eqref{eq:kgf} and use the approximate form of $\frac{q(\xi)}{q(\chi)}$ in Eq.~\eqref{eq:qsurq}. When $z$ lies below the real axis we thus have 
\begin{equation}\begin{split}
 \Gamma(z,z';t)\simeq &  \frac{3\ee^{-a^2 \sqrt{3}  t}}{2\beta N^2 a^2}
  \frac{1}{\sinh^2(\chi+\chi_1)}  \frac{1}{\sinh\chi\sinh\chi'
    \sinh^2\frac{\chi_1+\chi'}{2}}
    \end{split}
\end{equation}
When $z=\lambda- i0^+$ lies right below the real axis we can write that $\sinh\chi=-i\sin\theta$ with $\sin\theta=\sqrt{1-\lambda^2/a^2}$, which leads to
\begin{equation}\begin{split}
 \Gamma(\lambda-i0^+,z';t)\simeq &  \frac{3\ee^{-a^2 \sqrt{3}  t}}{2\beta N^2 a^2}
  \frac{1}{\sinh^2(-i\theta+\chi_1)}  \frac{1}{(-i)\sin\theta\sinh\chi'
    \sinh^2\frac{\chi_1+\chi'}{2}}
    \end{split}
\end{equation}
A similar formula can be obtained for $z=\lambda+i0^+$, paying attention to the fact that now the characteristic flows to $-z_1$:
\begin{equation}\begin{split}
 \Gamma(\lambda+i0^+,z';t)\simeq &  \frac{3\ee^{-a^2 \sqrt{3}  t}}{2\beta N^2 a^2}
  \frac{1}{\sinh^2(i\theta+\chi_1^*)}  \frac{1}{i\sin\theta\sinh\chi'
    \sinh^2\frac{\chi_1^*+\chi'}{2}}
    \end{split}
\end{equation}
We use the inversion formula in Eq.~\eqref{eq:PerronStieltjes} sequentially for $z$ (at fixed $z'$), namely we first determine
\begin{equation}\begin{split}
\frac{\Gamma(\lambda-i0^+,z';t)-\Gamma(\lambda+i0^+,z';t)}{2i\pi}\simeq&  \frac{3\ee^{-a^2 \sqrt{3} t}}{4\pi\beta N^2 a^2}
  \frac{1}{\sin\theta\sinh\chi'}\\
  &\times\left[\frac{1}{
    \sinh^2(-i\theta+\chi_1)\sinh^2\frac{\chi_1+\chi'}{2}}+\frac{1}{
    \sinh^2(i\theta+\chi_1^*)\sinh^2\frac{\chi_1^*+\chi'}{2}}\right]
\end{split}\end{equation}
and then we proceed similarly for $z'$ (upon setting $\sinh\chi'=\mp i\sin\theta'$ with $\sin\theta'=\sqrt{1-\lambda'^2/a^2}$ for $z'=\lambda'\mp i 0^+$). This eventually results in
\begin{equation}
\begin{split}
   C(\lambda,\lambda';t)\simeq &\frac{3\ee^{-a^2 \sqrt{3} t}}{8\pi^2\beta N^2 a^2}  \frac{1}{\sin\theta\sin\theta'}\\
   &+\left[\frac{1}{
    \sinh^2(-i\theta+\chi_1)\sinh^2\frac{\chi_1-i\theta'}{2}}+\frac{1}{
    \sinh^2(i\theta+\chi_1^*)\sinh^2\frac{\chi_1^*-i\theta'}{2}}\right.\\
    &\;\;\left.+\frac{1}{
    \sinh^2(-i\theta+\chi_1)\sinh^2\frac{\chi_1+i\theta'}{2}}+\frac{1}{
    \sinh^2(i\theta+\chi_1^*)\sinh^2\frac{\chi_1^*+i\theta'}{2}}\right]\\
    \simeq&\frac{12\ee^{-a^2 \sqrt{3} t}}{\pi^2\beta N^2 a^2}  \frac{1}{\sin\theta\sin\theta'}\\
    &\times\frac{1+2\cos(2\theta)+2\cos(2\theta')+4 \cos(2\theta)\cos(2\theta')-\sqrt{6}\cos\theta\sin\theta\cos\theta' (-5 + 
   2 \cos^2\theta')}{(2+\cos(2\theta))^2(2+\cos(2\theta'))^2}
    \end{split}
\end{equation}
which, in turn, simplifies into
\begin{equation}
    C(\lambda,\lambda';t)\simeq\frac{12a^3}{\beta N^2\pi^2 }\ee^{-a^2\sqrt{3}t}\frac{a(4\lambda^2-a^2)(4\lambda'^2-a^2)+\sqrt{6} \lambda\lambda'\sqrt{a^2-\lambda'^2}(5 a^2-2\lambda'^2)}{\sqrt{a^2-\lambda^2}\sqrt{a^2-\lambda'^2}(2\lambda^2+a^2)^2(2\lambda'^2+a^2)^2}
\end{equation}
The observation that at large times the correlations become attractive therefore survives in the case of a quartic potential for eigenvalues close to the origin.

\subsection{Beyond the quartic potential}
We finally consider a potential $V$ that is an even polynomial with degree $n>4$. Since $Q$ is an even polynomial of degree $n-2$ it can always be written as
\begin{equation}
    Q(z)=A\prod_{j=1}^{n/2-1}(z^2-z_j^2)
\end{equation}
where $A>0$ and the $n-2$ zeros $\pm z_j$ lie in the complex plane (away from the $[-a,a]$ branch cut). By convention we choose $z_j$ with a negative imaginary part. One of the difficulties is that, depending on the location of $z$ in the interval $[-a, a]$, the characteristic $\zeta(t)$ may flow to a finite number of distinct locations in the complex plane, which renders a fully general approach rather cumbersome. Indeed, using that
\begin{equation}
    \frac{1}{Q(z)}=\sum_j \frac{2 z_j}{Q'(z_j)}\frac{1}{z^2-z_j^2}
\end{equation}
we find the implicit solution of the characteristic to be given by
\begin{equation}\label{eq:t-asanintegral}
    t=\sum_j\frac{2z_j}{Q'(z_j)}\int_z^\zeta\dd \xi\frac{1}{ \sqrt{\xi^2-a^2}}\frac{1}{\xi^2-z_j^2}
\end{equation}
%If the starting point $z$ is real and outside the branch cut $[-a,a]$, we see that
%for $\zeta\to\infty$ each of the above integrals is finite. Therefore, unlike what is happening for a quadratic potential, it takes a finite time $\tau(z,\infty)$ for $\zeta$ to reach $\infty$. For $t>\tau(z,\infty)$, $\zeta(t)$ is finite again. 
%This only happens when the starting point $z\in \mathbb{R}\backslash[-a,a]$ is real. For $z$ not real, $\zeta(t)$ remains bounded at all times. 
When $t$ grows to infinity the $\zeta(t)$ trajectory in the complex plane must eventually approach one of the $\pm z_j$'s (while keeping $t$ a positive real number). This can be seen from the explicit integration of Eq.~\eqref{eq:t-asanintegral}, which leaves us with
\begin{equation}\label{eq:solcomp}
    t=\sum_j\frac{1}{Q'(z_j)\sqrt{z_j^2-a^2}}\ln\frac{\left[\sqrt{\zeta^2-a^2}z_j-\zeta\sqrt{z_j^2-a^2}\right]\left[\sqrt{z^2-a^2}z_j+z\sqrt{z_j^2-a^2}\right]}{\left[\sqrt{\zeta^2-a^2}z_j+\zeta\sqrt{z_j^2-a^2}\right]\left[\sqrt{z^2-a^2}z_j-z\sqrt{z_j^2-a^2}\right]} 
\end{equation}
Solving for $\zeta$ as a function of $t$ and of the starting point $z$ is out of analytical reach for an arbitrary potential. If the flow would bring $\zeta$ close to $z_j$ then $\zeta(t)-z_j$ would shrink exponentially to zero in $t$ with a rate $Q'(z_j)\sqrt{z_j^2-a^2}$ (which must thus have a negative real part). While for $t\to\infty$ we indeed expect that $\zeta$ approaches $\pm z_j$, it is unclear which $z_j$ is accessible from a given initial $z$. Keeping in mind that, ultimately we want to revert to eigenvalue space, which requires the use of the inversion formula in Eq.~\eqref{eq:PerronStieltjes}, we have to use as a starting point of the flow either $z=\lambda-i0^+$  or $z=\lambda+i0^+$, and $\lambda\in[-a,a]$. To be concrete, we choose the initial point to lie somewhere right below the $[-a,a]$ branch cut.\\

In the case of $\lambda^6$ polynomial, the phenomenology of the characteristics is very similar to that of a quartic potential, because then $Q$ possesses two roots $z_1$ and $z_2$ in the half-plane $\text{Im}(z_j)<0$, but $-z_2^*=z_1$, which both lead to identical values of $Q'(z_j)\sqrt{z_j^2-a^2}$ for $j=1$ or $j=2$. But the relaxation scenario becomes much richer for a polynomial of degree $8$ or higher.\\

\begin{figure}[h]
    \centering
   % \includegraphics[width=10cm]{flowoctic.pdf}
   %\begin{overpic}[width=10cm,grid,tics=10]{flowoctic.pdf}
   \begin{overpic}[width=10cm]{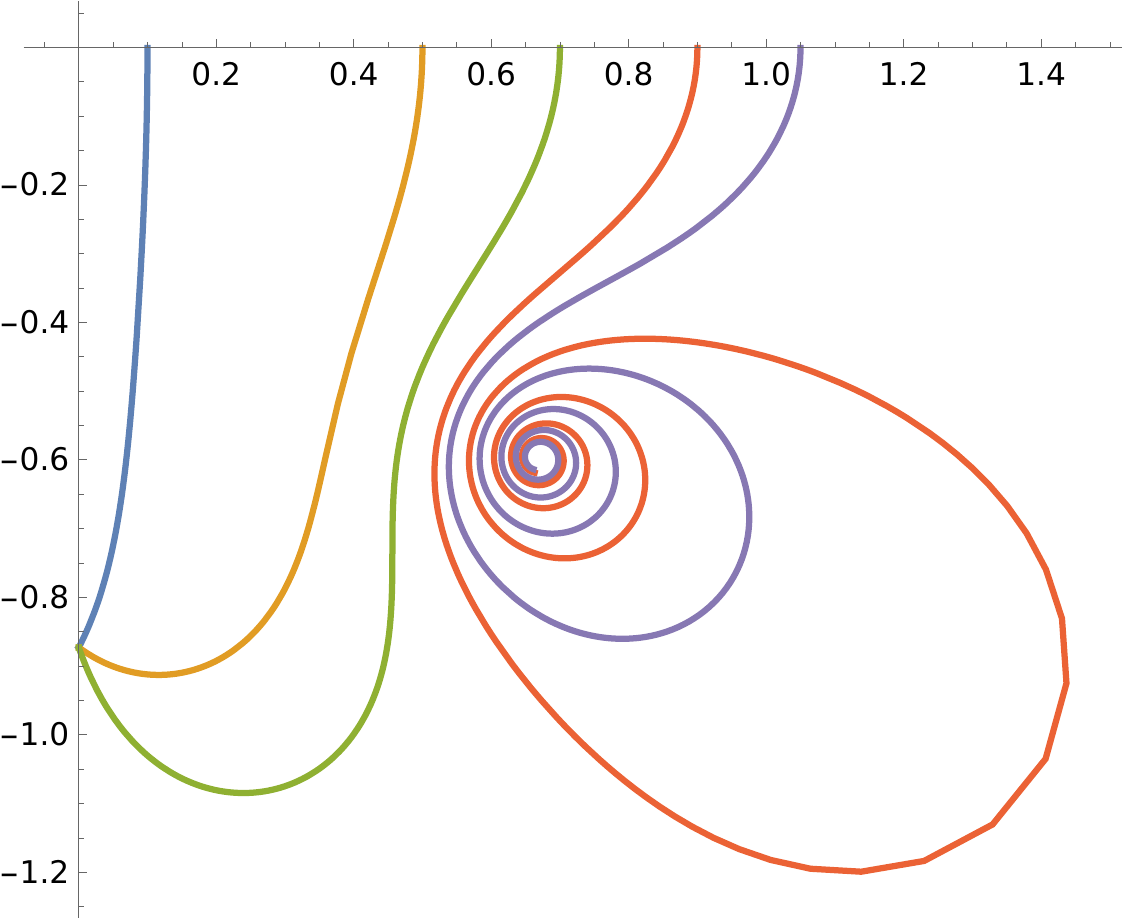}
 \put  (3,24) {$z_1$}
 \put  (72,28) {$z_2$}
 \put (6,83) {$\text{Im}(\zeta)$}
\put (102,77) {$\text{Re}(\zeta)$}
\linethickness{2pt}
\put(72,29){\color{black}\vector(-2,1){24}}
\end{overpic}

\caption{Trajectories of the characteristics starting from right below the branch cut $[-a,a]$ for a potential $V(\lambda)=\frac{\lambda^8}{8}$ ($Q(\lambda)=\lambda^6+\frac{a^2}{2}\lambda^4+\frac{3}{8}a^4 \lambda^2+\frac{5}{16}a^6$ and $a^8=64/35$). The roots are $z_1\simeq-0.9 i$, $z_2=0.7-0.6 i$ (and $z_3=-z_2^*$) and $a\simeq 1.1$. The characteristics are initialized with positive reals parts (due to the symmetries of the potential) at the following (just below the real axis): $z=0.1$ (blue), $z=0.5$ (orange), $z=0.7$ (green), $z=0.9$ (red) and $z=1.05$ (violet). The basins of attraction of each $z_1$ and $z_2$ are non trivial.}\label{fig:flotoctic}
\end{figure}
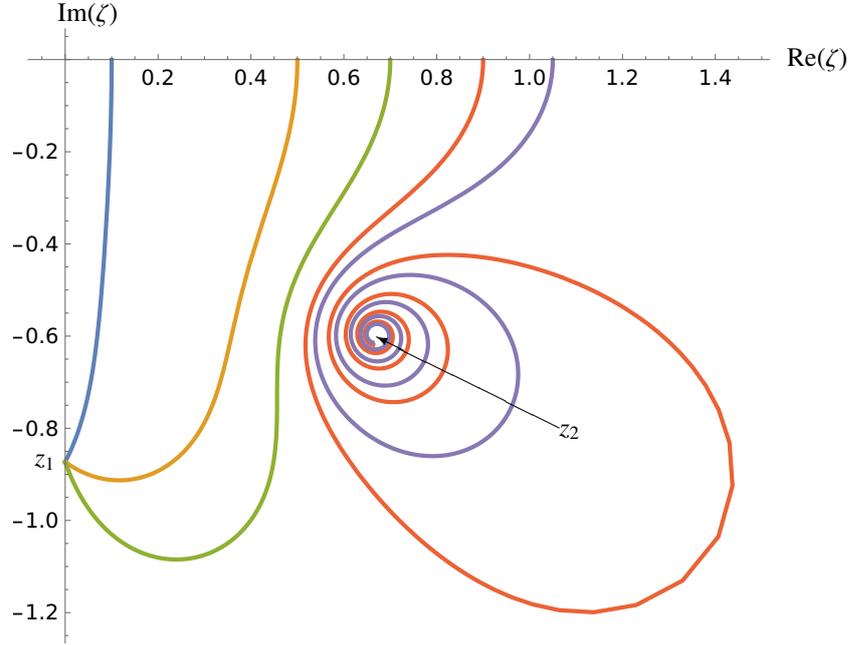

Indeed, for $V(\lambda)=\frac{\lambda^8}{8}$, the corresponding $Q$ polynomial is of degree $6$, and there are thus three roots $z_1$, $z_2$ and $z_3$ of interest. It turns out that $z_1\in i\mathbb{R}$ while $z_3=-z_2^*$. In practice this means that we find two distinct relaxation scales $Q'(z_1)\sqrt{z_1^2-a^2}\simeq -3.7$ and $Q'(z_2)\sqrt{z_2^2-a^2}\simeq -1.4+4.2 i$. As can be seen in Fig.~\ref{fig:flotoctic}, the approach of $z_2$ comes along with a spiral motion due to the existence of an imaginary part to $Q'(z_2)\sqrt{z_2^2-a^2}$. This translates into damped oscillations for the time decay of the correlations.
It is a remarkable feature that the relaxation rate of density correlations depends on the eigenvalues that are being considered. This feature is expected to hold for polynomials of degree $8$ and higher.

\section{What's next}
To conclude this work we would like to list what seem to us to be promising research directions by increasing order of speculation.
To begin with, at $\beta=2$ the dynamics is exactly solvable because the evolution operator reduces to that of a set of free fermions. It would be interesting to recover our expression by a direct method (which, unlike ours, would be valid at arbitrary $N$). It will undoubtedly involve the same family of orthogonal polynomials considered by Brézin and Zee~\cite{brezin1993universality}, but it is unclear how the characteristics will emerge. These characteristics and their complexity are responsible for the emergence of several time scales. Understanding this phenomenon for an arbitrary potential is certainly worthy of interest.\\

Second, in the absence of a confining potential, there is no equilibrium since the eigenvalues flee away at infinity under the influence of the repulsive Coulomb force. This can nicely be captured by means of the resolvent~\cite{warchol2014dynamic}. Denoting by $G_\text{t}(z,t)$ the mean  resolvent at time $t$ we have the expected inviscid and noisy Burgers equation. Given the initial state $G_\text{t}(z,0)=G_0(z)$ it is possible~\cite{bun2014instanton} to find the time-dependent solution by the method of characteristics ($G_\text{t}(z,t)=G_0\left(z- G_\text{t}(z,t) t\right)$) so that the mean-behavior is well under control. For instance, preparing the system with a harmonic well at $t=-\infty$ and removing the well at time $t=0$ leads to the relaxation $G_\text{t}(z,t)=\frac{z-\sqrt{z^2-2-4t}}{1+2t}$, which expresses the diffusive spreading of the initial semi-circle. If on the contrary all eigenvalues are lumped at the origin, then $G_\text{ts}(z,t)=\frac{z-\sqrt{z^2-4t}}{2t}$. In principle what our formalism now allows us to access are the aging and relaxation properties (the corresponding characteristics $\zeta(t)$ are of course more involved because of the explicit time dependence in the analog of Eq.~\eqref{eq:charact}).\\

Third, the study of large deviations in random matrices has a very rich recent history~\cite{dean2006large,majumdar2009index,majumdar2012number,majumdar2014top}. The common thread to these works is to exploit the thermodynamic formulation based on the functional $E[\hat{\mu}]$ of Eq.~\eqref{eq:energy} and variations around it. A direct dynamical extension of this thermodynamic methods is the recent work on current fluctuations (with very weak confinement so as to preserve translation invariance) of \cite{kirone2024}. It would be interesting to see the extent to which resolvent-based methods can be useful in probing dynamical large deviations, whether of global quantities or of extreme eigenvalues (or of a tagged eigenvalue). Incidentally this need not be limited to the three ensembles studied in this work; probing Wishart matrices and the Ginibre ensemble (building on \cite{burda2014dysonian}) seems within reach.\\

Finally, as has appeared in this work, being able to work with  the resolvent within a weak noise approximation has been central  in our approach. As is well-known, the resolvent is also instrumental in building the so-called ${\mathcal R}$-transform that appears in free probability~\cite{voiculescu1997free,voiculescu1992free,bousseyroux2024free}. This function  $\hat{\mathcal{R}}(z,t)$ is defined by $\hat{G}(\hat{\mathcal R}(z,t)+z^{-1},t)=z$. It is thus a fluctuating and time-dependent quantity. Whether our equation for $\hat{G}$ translates into something fruitful for the dynamical $\hat{\mathcal R}$ remains to be explored. 

\section*{Acknowledgements}
We acknowledge the financial support of the ECOS Nord C24P01 joint project and Minciencias, Patrimonio Autónomo Fondo Nacional de
Financiamiento para la Ciencia, la Tecnología y la Innovación, Francisco José de
Caldas. FvW and GT thank Abhishek Dhar, Hisao Hayakawa and Peter Forrester for useful feedback. GT acknowledges support of Universidad de los Andes, Facultad de
Ciencias, project number INV-2023-176-2951.

\appendix

\section{Equal time correlations of the resolvent}\label{app:corr}
In this appendix, we establish the expression of Eq.~\eqref{eq:correl-resolv-eq} for $\Gamma_\text{eq}(z,z')$. We start from the known expression found in \cite{brezin1993universality,BEENAKKER1994515} for $C_\text{eq}(\lambda,\lambda')$ given in Eq.~\eqref{eq:brezinzee}, which is valid for $\lambda\neq\lambda'$, and that we recall below for convenience:
\begin{equation}
C_\text{eq}(\lambda,\lambda')=-\frac{1}{\beta N^2\pi^2}\frac{a^2-\lambda\lambda'}{\sqrt{a^2-\lambda^2}\sqrt{a^2-\lambda'^2}(\lambda-\lambda')^2}
\end{equation}
For $\lambda=\lambda'$ there is a subleading (in $N$) contribution that we shall ignore, and we thus set $C_\text{eq}(\lambda,\lambda)=0$. We are interested in determining the expression of its double Stieltjes transform
\begin{equation}
    \Gamma_\text{eq}(z,z')=\int_{-a}^a\frac{\dd\lambda}{z-\lambda}\int_{-a}^a\frac{\dd\lambda'}{z'-\lambda'}C_\text{eq}(\lambda,\lambda')
\end{equation}
which we consider for  $z\neq z'$, both $z$ and $z'$ being outside the branch cut defined by the $\sqrt{a^2-\lambda^2}$ and $\sqrt{a^2-\lambda'^2}$ contributions. We begin by introducing the two angles $\theta$ and $\theta'$ in $[0,\pi]$ defined by
\begin{equation}
    \cos\theta=\frac{\lambda}{a},\,\,\cos \theta'=\frac{\lambda'}{a}
\end{equation}
and the corresponding angle-like variables $\chi$ and $\chi'$ for $z$ and $z'$
\begin{equation}
    \cosh\chi=z/a,\,\,\cosh\chi'=z'/a
\end{equation}
In terms of these new variables the double Stieltjes transform reads
\begin{equation}\begin{split}
    \Gamma_\text{eq}(z,z')=&-\frac{1}{\beta N^2\pi^2a^2}\int_{0}^\pi\dd\theta\dd\theta' \frac{1-\cos\theta\cos\theta'}{(\cosh\chi-\cos\theta)(\cosh\chi'-\cos\theta')(\cos\theta-\cos\theta')^2}\\
    =&-\frac{1}{4\beta N^2\pi^2a^2}\int_{0}^{2\pi}\dd\theta\dd\theta' \frac{1-\cos\theta\cos\theta'}{(\cosh\chi-\cos\theta)(\cosh\chi'-\cos\theta')(\cos\theta-\cos\theta')^2}
    \end{split}
\end{equation}
where now $u=\ee^{i\theta}$ and $u'=\ee^{i\theta'}$ run along the unit circle around the origin. In terms of these new variables we thus have to determine
\begin{equation}
    \Gamma_\text{eq}(z,z')=-\frac{1}{\beta N^2\pi^2a^2}\oint\dd u\dd u'\underbrace{ \frac{u u'((u^2+1)(u'^2+1)-4 u u')}{(u-u')^2(u u'-1)^2(u-\ee^\chi)(u-\ee^{-\chi})(u'-\ee^{\chi'})(u'-\ee^{-\chi'})}}_{h(u')}
\end{equation}
For concreteness we choose to integrate first along the $u'$ unit circle. In order to avoid any singularity we shall shift the radius of the $u$ integration by an infinitesimal quantity, so that $u$ now lies slightly within or slightly beyond the unit circle (depending on the sign of the shift). As a function of $u'$ the integrand $h(u')$ has only isolated singularities, namely a pair of  poles of order 2 located at $u'=u$ and $u'=1/u$, and a pair of poles of order 1 located at $u'=\ee^{\chi'}$ and $u'=\ee^{-\chi'}$. We can perform the calculation because we have chosen to work with $u$ not sitting on the unit circle (and $\ee^{\chi'}$ does not sit on the unit circle either). This requires determining only two residues: if $u$ is inside the unit circle then $1/u$ is outside and it does not contribute the contour integral, and the same applies to $\ee^{\chi'}$. An explicit calculation shows that
\begin{equation}\begin{split}
    \text{Res}_u(h)=-\text{Res}_{1/u}(h)=\frac{ u (1-u^2)}{(u-\ee^\chi)(u-\ee^{-\chi})(u-\ee^{\chi'})^2(u-\ee^{-\chi'})^2} \\
    \text{Res}_{\ee^{\chi'}}(h)=-\text{Res}_{\ee^{-\chi'}}(h)=\frac{1}{\sinh\chi'}\frac{u((u^2+1)\cosh\chi'-2u)}{(u-\ee^\chi)(u-\ee^{-\chi})(u-\ee^{\chi'})^2(u-\ee^{-\chi'})^2}
\end{split}\end{equation}
and thus by the residue theorem (say for $\chi'>0$ and, say, for $u$ just inside the unit circle) we arrive at
\begin{equation}\begin{split}
    \Gamma_\text{eq}(z,z')=&-\frac{2\pi i}{\beta N^2\pi^2a^2}\oint\dd u\left[\frac{ u (1-u^2)}{(u-\ee^\chi)(u-\ee^{-\chi})(u-\ee^{\chi'})^2(u-\ee^{-\chi'})^2}\right.\\
    &-
    \left.\frac{1}{\sinh\chi'}\frac{u((u^2+1)\cosh\chi'-2u)}{(u-\ee^\chi)(u-\ee^{-\chi})(u-\ee^{\chi'})^2(u-\ee^{-\chi'})^2}
    \right]\\
    =&-\frac{2\pi i}{\beta N^2\pi^2a^2}\oint\dd u\frac{-u (1+\coth\chi')}{2(\ee^{\chi'}-u)^2(u-\ee^\chi)(u-\ee^{-\chi})}
\end{split}\end{equation}
And again we are left with an integral of a function of $u$ with one pole of order 2 located at $\ee^{\chi'}$ (outside of the unit circle), and two poles of order 1 located at $\ee^{\pm\chi}$. For $\chi>0$ a single pole contributes and we directly arrive at
\begin{equation}\begin{split}
    \Gamma_\text{eq}(z,z')=&-\frac{(2\pi i)^2}{\beta N^2\pi^2a^2}\frac{1}{8 \sinh\chi\sinh\chi'\sinh^2\frac{\chi+\chi'}{2}}\\
    =&-\frac{1}{\beta N^2}\frac{a^2-z z'+\sqrt{z^2-a^2}\sqrt{z'^2-a^2}}{\sqrt{z^2-a^2}\sqrt{z'^2-a^2}(z-z')^2}
\end{split}\end{equation}
Note that if we had chosen a regularization where $u$ lies slightly outside the unit circle, we would have ended up on an integrand involving a pole of order 1  at $\ee^{-\chi}$ and one of order 2 at $\ee^{-\chi'}$, with exactly the same final result (this final result does not depend on our choice of regularization). The expression is analytic in $\chi$ and $\chi'$ so that the expression extends over the whole range of $\chi$ and $\chi'$. This establishes the expression of $\Gamma_\text{eq}$ used in 
Eq.~\eqref{eq:correl-resolv-eq}.
\section{Some side technical comments}\label{subsec:technical}
In our derivation we have relied on the knowledge of equilibrium correlations (we use $\Gamma_\text{eq}$ as an input). But in principle our approach need not rest on this {\it a priori} knowledge. The Dyson Brownian is self-contained and it knows about the equilibrium state. Let us briefly sketch how we could circumvent this {\it a priori} knowledge. As we have seen, $R$ is the Green's function of a differential operator (as is seen in Eq.~\eqref{eq:forRgeneral}) and it is therefore tempting to use it to extract $\Gamma$ by convoluting $R$ with Eq.~\eqref{eq:forGamma}:
\begin{equation}
    \Gamma(z,z';t-t')=-\frac{2}{\beta N^2}\int_{-\infty}^{+\infty}\dd s \dd z_1\dd z_2 R(z,z_1;t-s)\p_{z_1}\p_{z_2}\frac{G_\sW(z_1)-G_\sW(z_2)}{z_1-z_2}R(z',z_2;t'-s)
    \end{equation}
Using the explicit form of $R$ leads to
\begin{equation}\begin{split}
    \Gamma(z,z';t-t')=&-\frac{2}{\beta N^2}\frac{1}{Q(z)\sqrt{z^2-a^2}Q(z')\sqrt{z'^2-a^2}}\\
    &\times\int_{-\infty}^{\min\{t,t'\}}\dd s Q(\zeta)\sqrt{\zeta^2-a^2}Q(\zeta')\sqrt{\zeta'^2-a^2}\p_{\zeta}\p_{\zeta'}\frac{G_\sW(\zeta)-G_\sW(\zeta')}{\zeta-\zeta'}
\end{split}\end{equation}
where $\zeta=\zeta(t-s)$ with $\zeta(0)=z$, and $\zeta'=\zeta(t'-s)$ with $\zeta'(0)=z'$. We remark that
\begin{equation}\label{eq:conjGamma}\begin{split}
    \Gamma(z,z';t-t')=&-\frac{1}{\beta N^2}\frac{1}{Q(z)\sqrt{z^2-a^2}Q(z')\sqrt{z'^2-a^2}}\\&\times\Bigg[\int_{-\infty}^{\min\{t,t'\}}\dd s\frac{\dd}{\dd s} \left( Q(\zeta)Q(\zeta')\frac{a^2-\zeta\zeta'+\sqrt{\zeta^2-a^2}\sqrt{\zeta'^2-a^2}}{(\zeta-\zeta')^2}\right)\\
    &+\int_{-\infty}^{\min\{t,t'\}}\dd s f(\zeta,\zeta')\Bigg]
\end{split}\end{equation}
where $f$ is a function that is regular as $\zeta\to\zeta'$. In the particular case of a quartic potential it takes the form
\begin{equation}
    f(\zeta,\zeta')= Q(\zeta)Q(\zeta')\sqrt{\zeta^2-a^2}\sqrt{\zeta'^2-a^2}\left(\frac{\zeta}{\sqrt{\zeta^2-a^2}}+\frac{\zeta'}{\sqrt{\zeta'^2-a^2}}-2\right)
\end{equation}
and one quickly realizes that the $s$ integral of $f$ diverges. The reason for the failure of this direct but too naive approach as we just presented it lies in a mathematical subtlety. While $R$ is the Green's function of a linear first order (in time and $z$) operator, it can only be used to solve the partial differential equation for $\Gamma(z,z',t)$ if the source term vanishes sufficiently fast at infinity. When this is not the case, the extra terms involving the behavior at infinity must be taken into account. This leads to counterterms that leave us with the first line of Eq.~\eqref{eq:conjGamma} and this suppresses the contribution from $f$. Given we have an alternative path to derive the correlations, we do not pursue this line of reasoning any further.

\bibliographystyle{elsarticle-num}
\bibliography{biblio-DBM}

\end{document}